\documentclass[aps,preprint,showpacs,nofootinbib,12pt]{revtex4}
\usepackage{graphicx}
\usepackage{epsfig}
\usepackage[dvips]{color}

\begin{document}

\title{ The study of possible molecular states of $D_s^{(*)}D_s^{(*)}$ and $B_s^{(*)}B_s^{(*)}$}

\author{ Hong-Wei Ke$^{1}$   \footnote{khw020056@tju.edu.cn} and Yan-Liang Shi
        $^2$\footnote{shi@cshl.edu} }

\affiliation{
  $^{1}$ School of Science, Tianjin University, Tianjin 300072, China \\
  $^{2}$ Cold Spring Harbor Laboratory, Cold Spring Harbor, NY 11724, USA
 }

\begin{abstract}
\noindent

Recently the LHCb collaboration reported a new exotic state $T^+_{cc}$ which
is conjectured to be a molecular state
of $D^0D^{*+}$ (or $D^{*0}D^{+}$) theoretically.  Belle Collaboration also searched for tetraquark state $X_{ccss}$ in $D_sD_s (D^*_sD^*_s )$ final states but no significant signals were observed, which did not rule out the possibility of $X_{ccss}$ to be a molecular state of $D_sD_s (D^*_sD^*_s )$. Inspired by these experimental results on double charmed exotic state, in this paper we study whether the molecular bound states of $D_s^{(*)}D_s^{(*)}$ and $B_s^{(*)}B_s^{(*)}$ can exist with the Bethe-Salpeter (B-S) equation approach. We employ heavy meson chiral perturbation theory and one-boson-exchange approximation to calculate  the interaction kernels in the B-S
equations. Our numerical results suggest that  two $B_s^{*}$ mesons perhaps form a $0^+$ molecular state. Future experimental search for $X_{ccss}$ states in other decay  channels  may shed light on the structure of double charmed exotic state.

\end{abstract}


\maketitle

\section{introduction}
Several months ago the LHCb Collaboration declared a new exotic state $T_{cc}^+$
from the $D^0D^0\pi^+$ final state, which indicates
$T_{cc}^+$ possesses a $cc\bar u \bar d$ flavor component. Since  its mass is very close to the mass threshold of $D^0D^{*+}$  and its width is
very nerrow\cite{LHCb:2021auc,LHCb:2021vvq},  many authors suggested
that $T_{cc}^+$ could be a loose $D^0D^{*+}$ ($D^+D^{*0}$) bound
state\cite{Meng:2021jnw,Yan:2021wdl,Ren:2021dsi,Du:2021zzh,Xin:2021wcr,Feijoo:2021ppq,Dai:2021vgf,Deng:2021gnb,Zhao:2021cvg,Chen:2021cfl,Agaev:2021vur}. Since 2003, many
exotic
states\cite{Choi:2003ue,Abe:2007jn,Choi:2005,Choi:2007wga,LHCb:2021uow,Collaboration:2011gj,LHCb:2019kea} have been observed,
such as $X(3872)$, $X(3940)$,
$Y(3940)$, $Z(4430)^{\pm}$, $Z_{cs}(4000)$, $Z_{cs}(4220)$, $Z_b$,
$Z_b'$, $P_c(4312)$, $P_c(4440)$, $P_c(4457)$ but heavy quarks in them are hidden. If $T_{cc}^+$ is confirmed, it will be the first exotic state with two open heavy quarks.

Given a potential $T_{cc}^+$ state,  naturally one would ask whether the exotic states with $cc\bar s \bar s$ flavor component can also exist. Some theoretical works have explored these exotic states\cite{Li:2012ss,Liu:2019stu,Dai:2022ulk,Ding:2021igr}.  Recently Belle Collaboration searched for tetraquark state $X_{ccss}$ in $D_sD_s (D^*_sD^*_s )$ final states but no significant signals were observed\cite{Belle:2021kub}. However,  we cannot rule out the possible  $D_sD_s (D^*_sD^*_s )$ molecular state of  $X_{ccss}$  from the Belle's experiment because a ground molecular state cannot decay to its two components.

In a recent paper\cite{Ke:2021rxd} we study the possible
bound states of $D^0D^+$, $D^0D^{*+}$ and $D^{*0}D^{*+}$ ($B^0B^+$, $B^0B^{*+}$ and $B^{*0}B^{*+}$ ) systems
within the Bethe-Salpeter (B-S) framework where the relativistic corrections are automatically included. In this work we follow the same approach to explore the possible bound state of $D_sD_s$, $D_sD^{*}_s$ or $D^{*}_sD^{*}_s$ ($B_sB_s$, $B_sB^{*}_s$ or $B^{*}_sB^{*}_s$) system.

Apart form the bound state of two fermions, the B-S equation has also been employed to explore
the bound state made of one fermion and one
boson\cite{Guo:1998ef,Weng:2010rb,Li:2019ekr}, and the system composed of two bosons\cite{Guo:2007mm,Feng:2011zzb,Ke:2018jql,Feng:2012zzf,Ke:2012gm,Ke:2020eba,Ke:2019bkf,Ke:2021iyh}.
In Refs. \cite{Guo:2007mm,Feng:2011zzb,Ke:2012gm,Ke:2020eba,Ke:2019bkf} two components in the bound state are one particle and one
	antiparticle. In Ref. \cite{Ke:2021rxd}  the systems composed of two charmed (or bottomed)
hadrons are studied. Following the approach in Ref. \cite{Ke:2021rxd} we investigate the possible bound state of $D^{(*)}_sD^{(*)}_s$ ($B^{(*)}_sB^{(*)}_s$). 

 In this work we use the heavy meson chiral
 perturbation
 theory\cite{Colangelo:2005gb,Colangelo:2012xi,Ding:2008gr,Casalbuoni:1996pg,Casalbuoni:1992gi,Casalbuoni:1992dx} to describe the interaction between constituent mesons. Then we apply one-boson-exchange approximation of effective interaction to calculate the interaction kernels for the B-S equation. For  $D_s^{(*)}D_s^{(*)}$ and $B_s^{(*)}B_s^{(*)}$ systems, the exchanged particles are light mesons, such as $\eta$ and $\phi$. We ignore the contribution from
 $\sigma$ exchange because  some
authors indicated that  makes a secondary
contribution\cite{Ding:2008gr}. By calculating the corresponding Feymann diagrams for the effective interaction,  one can obtain the
analytical form of interaction kernel and deduce the B-S equation.
With  input parameters, the B-S equation is solved
numerically in momentum space. In the case where we cannot find solution that satisfies the
equation given a reasonable range of parameters, the proposed bound state cannot exist. On the contrary,  a solution of the B-S equation with reasonable
parameters implies that the interaction between two constituents is attractive and
large enough i.e. the corresponding bound state could be formed.

After introduction we deduce the B-S equations and the corresponding kernels for the $D^0D^+$, $D^0D^{*+}$ and $D^{*0}D^{*+}$ ($B^0B^+$, $B^0B^{*+}$ and $B^{*0}B^{*+}$ )  systems with defined quantum numbers. Then in section III we
present our numerical results along with
explicitly displaying all input parameters. Section IV is devoted
to a brief summary.

\section{The Bethe-Salpeter formalism}

In this work we are only concerned with the ground state where the orbital
angular momentum between two constituent mesons is zero (i.e. $l=0$).
For a system made of $D_s$ and $D^*_s$ (or $B_s$ and $B^*_s$), its $J^{P}$ is $1^+$. For
the molecular state which consist of $D_s$ and $D_s$ ( $B_s$ and $B_s$) their
$J^{P}$ may be $0^+$. However for
the molecular state which consists of  $D^*_s$ and $D^*_s$ ( $B^*_s$ and $B^*_s$) their
$J^{P}$ may be $0^+$ or $2^+$ rather than $1^+$ because the total wave function for the combined
system of   $D^*_s$ and $D^*_s$ ( $B^*_s$ and $B^*_s$) must be
symmetric under group $O(3)\times SU_S(2)$, where
 $SU_S(2)$ is the spin group.

\subsection{The B-S equation of $0^+$ which is composed of two pseudoscalars}

The B-S wave function for the bound state $|S\rangle$ of two
pseudoscalar  mesons can be defined  as following:
\begin{eqnarray}\label{definition-BS1} \langle 0 | {\rm
T}\,\phi_1(x_1)\phi_2(x_2) | S \rangle = {\chi}_{{}_S}^{}(x_1,x_2)\,,
\end{eqnarray}
where $\phi_1(x_1)$ and $\phi_2(x_2)$ are the field operators of two
mesons, respectively.

After some manipulations we obtain the B-S equation in the momentum space
\begin{eqnarray}\label{bs-equation-momentum1}
{\chi}_{{}_S}^{}(p) = \Delta_1\int {d^4p'\over
(2\pi)^4} { K_{S}}(p,p'){\chi}_{{}_S}^{}(p')\Delta_{2}\,,
 \end{eqnarray}
where $ \Delta_1=\frac{i}{p_1^2-m_1^2}$ and $ \Delta_2=\frac{i}{p_2^2-m_2^2}$ are the propagators of two pseudoscalar mesons.

The relative momenta and the total
momentum of the bound state in the equation are defined as
\begin{eqnarray} p = \eta_2p_1 -
\eta_1p_2\,,\quad p' = \eta_2p'_1 - \eta_1p'_2\,,\quad P = p_1 +
p_2 = p'_1 + p'_2 \,, \label{momentum-transform1}
\end{eqnarray}
where $\eta_i = m_i/(m_1+m_2)$, $P$ denotes the total momentum of the bound
state and $m_i\,
(i=1,2)$ is the mass of the $i$-th constituent meson.

Since only $l=0$ is considered and the total wavefunction of $D_sD_s$ is
symmetric i.e. the $J^{P}$ of  $D_sD_s$ system is  $0^+$. According to heavy meson chiral perturbation theory \cite{Colangelo:2005gb,Colangelo:2012xi,Ding:2008gr,Casalbuoni:1996pg,Casalbuoni:1992gi,Casalbuoni:1992dx}, the exchanged mesons
between the two pseudoscalars are vector mesons, and here we only keep the lightest vector meson $\phi$ \cite{Guo:2007mm,Feng:2011zzb}.
\begin{center}
\begin{figure}[htb]
\begin{tabular}{cc}
\scalebox{0.6}{\includegraphics{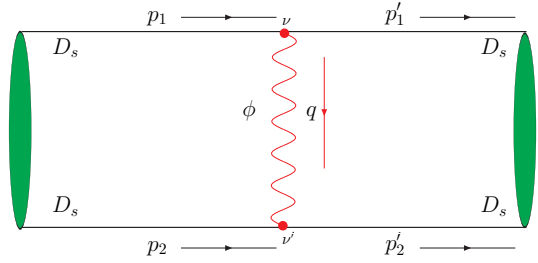}}
\end{tabular}
\caption{ The leading order interaction diagram of a bound state composed of two pseudoscalars.
}\label{DM21}
\end{figure}
\end{center}

With the Feynman diagrams depicted in Fig. \ref{DM21} and the effective
interactions shown in appendix A we obtain the interaction kernel
\begin{eqnarray}
&&K_{S}(p,p')=\sqrt{2}K_{S0}(p,p',m_\phi),\nonumber\\&&{K_{S0}}(p,p',m_{V}) = i
 C_{S0} \, g_{DDV}^2
{(p_1+p_1')\cdot(p_2+p_2')-(p_1+p_1')\cdot q(p_2+p_2')\cdot q/m_{
V}^2 \over q^2-m_{ V}^2}F( q)^2,
\end{eqnarray}
where $q=p_1-p_1'$, $C_{S0}=1$, and $m_V$ stands for mass of vector mesons. Because two $D_s$ mesons are identical particles, $p'_1$ and $p'_2$ can exchange positions in Fig. \ref{DM21} and an additional $\sqrt{2}$ factor in the right side of the first equation (a similar factor also applies to the  $D^*_sD^*_s$ systems in Section. \ref{two_v}). Since the constituent meson is not a point particle,
a form factor at each interaction vertex among hadrons must be
introduced to reflect the finite-size effects of these hadrons.
The form factor is assumed to be in the following form:
\begin{eqnarray} \label{form-factor} F({k}) = {\Lambda^2 -
M_{ex}^2 \over \Lambda^2 -{k}^2}\,,
\end{eqnarray} where $\Lambda$ is a cutoff parameter, and $M_{ex}$ is the mass of intermediate particle.

Solving the Eq. (\ref{bs-equation-momentum1}) is
rather difficult. In general one needs to use the so-called
instantaneous approximation: $p_0'=p_0=0$ for ${K_{0}}(p,p')$  by which
the B-S equation can be reduced to
\begin{eqnarray} \label{3-dim-BS1}
{E^2-(E_1+E_2)^2\over (E_1+E_2)/E_1E_2}
\mathcal{\psi}_{{}_S}^{}({\bf p}) ={i\over
2}\int{d^3\mathbf{p}'\over(2\pi)^3}\, {\overline{} K_{S}}({\bf
p},{\bf p}')\mathcal{\psi}_{{}_S}^{}({\bf p}')\,,
\end{eqnarray}
where $E_i \equiv \sqrt{{\bf p}^2 + m_i^2}$, $E=P^0$, and the
equal-time wave function is defined as $ \psi_{_S}({\bf p})= \int
dp^0 \, \chi_{_S}(p) \,. $ For  exchange of a light vector between
the mesons, the kernel is
\begin{eqnarray}K_{S}({\bf p},{\bf
p}')=\sqrt{2}K_{S0}({\bf p},{\bf p}',m_\phi),\end{eqnarray}
where the expressions of $K_{S0}(\mathbf{p},\mathbf{p}',m_V)$ can be found  in Ref.\cite{Ke:2021rxd}.

\subsection{The B-S equation of $1^+$ which is composed of a pseudoscalar and a vector}
The B-S wave function for the bound state $|V\rangle$ composed of
one pseudoscalar and one vector meson is defined as following:
\begin{eqnarray}\label{definition-BS2} \langle 0 | {\rm
T}\,\phi_1(x_1)\phi^\mu_2(x_2) | V \rangle ={\chi}_{{}_V}^{}(x_1,x_2)\epsilon^\mu\,,
\end{eqnarray}
where $\epsilon$ is the polarization vector of the bound state, $\mu$ is Lorentz index, $\phi_1(x_1)$ and $\phi^{\mu}_2(x_2)$ are the
field operators of the pseudoscalar and vector mesons, respectively. The equation for the B-S wave function is
\begin{eqnarray}\label{bs-equation-momentum2}
{\chi_{_V}}(p) \epsilon^{\mu} = \Delta_1 \int {d^4p'\over (2\pi)^4} {
K_{V\alpha\beta}}({ p},{
p}'){\chi_{_V}}(p') \epsilon^{\beta} \Delta_{2\mu\alpha}\,.
\end{eqnarray}
Here $ \Delta_1=\frac{i}{p_1^2-m_1^2}$ and $
\Delta_{2\mu\alpha}=\frac{i}{p_2^2-m_2^2}(\frac{p_{2\mu}p_{2\alpha}}{m_2^2}-g_{\mu\alpha})$. We
multiply an $\epsilon^*_\mu$ on both sides and sum over the
polarizations, and then we deduce a the following equation
\begin{eqnarray}\label{bs-equation-momentum2pp}
{\chi}_{{}_{V}}^{}(p) = \frac{-1}{3(p_1^2-m_1^2)(p_2^2-m_2^2)}\int
{d^4p'\over (2\pi)^4} {
K_{V\alpha\beta}}(p,p'){\chi}_{{}_{V}}^{}(p')(\frac{p_{2}^\mu
p_{2}^\alpha}{m_2^2}-g^{\mu\alpha})(\frac{P_\mu
P^\beta}{M^2}-g_\mu^\beta)\,.
 \end{eqnarray}

\begin{center}
\begin{figure}[htb]
\begin{tabular}{cc}
\scalebox{0.5}{\includegraphics{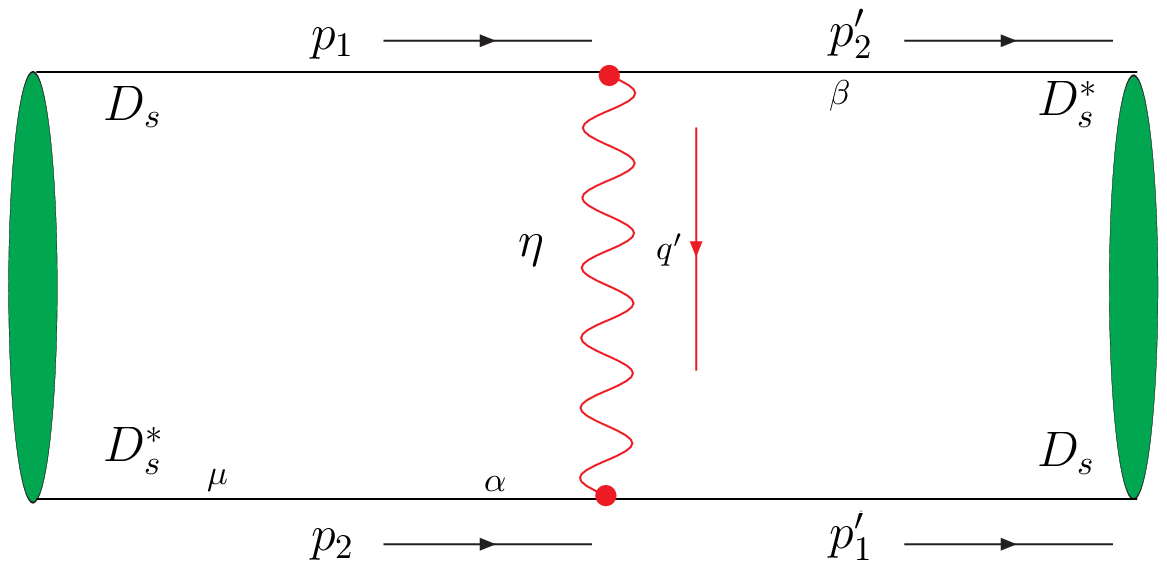}}
\end{tabular}
\caption{ The leading order interaction diagram of a bound state composed of a pseudoscalar and a vector by
exchanging $\eta$.}\label{DM22}
\end{figure}
\end{center}

\begin{center}
\begin{figure}[htb]
\begin{tabular}{cc}
\scalebox{0.5}{\includegraphics{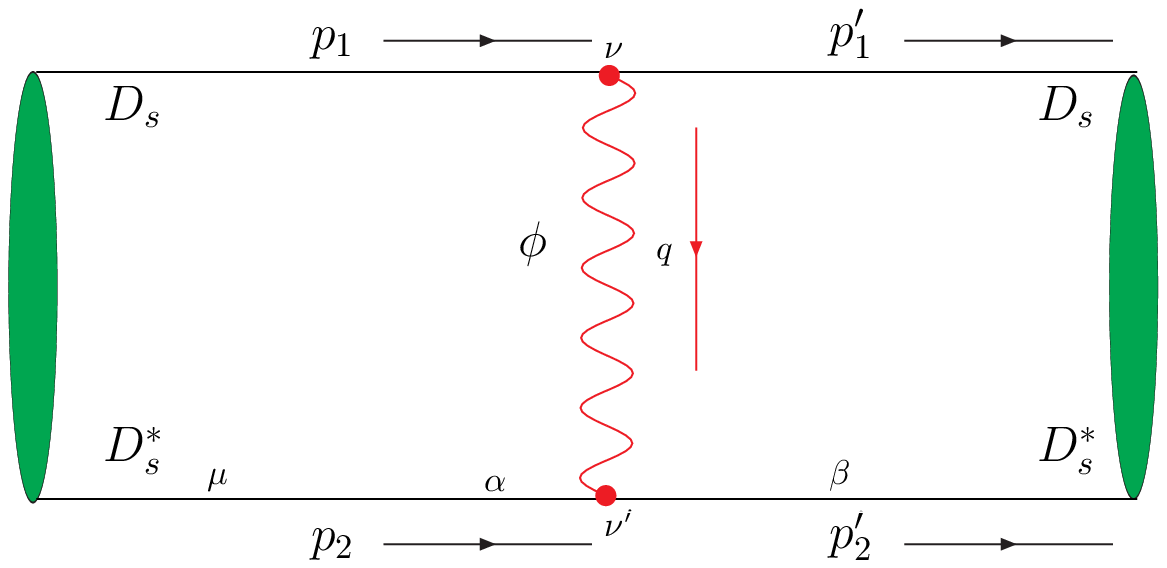}}\,\,\,\,\,\,\,\,\,\,\,\,\,\,\,\,\,\,\,\,\,\,\,\,\,\,\,\,\,\,\,\,\,\,\,\,
\,\,\,\,\,\,\,\scalebox{0.5}{\includegraphics{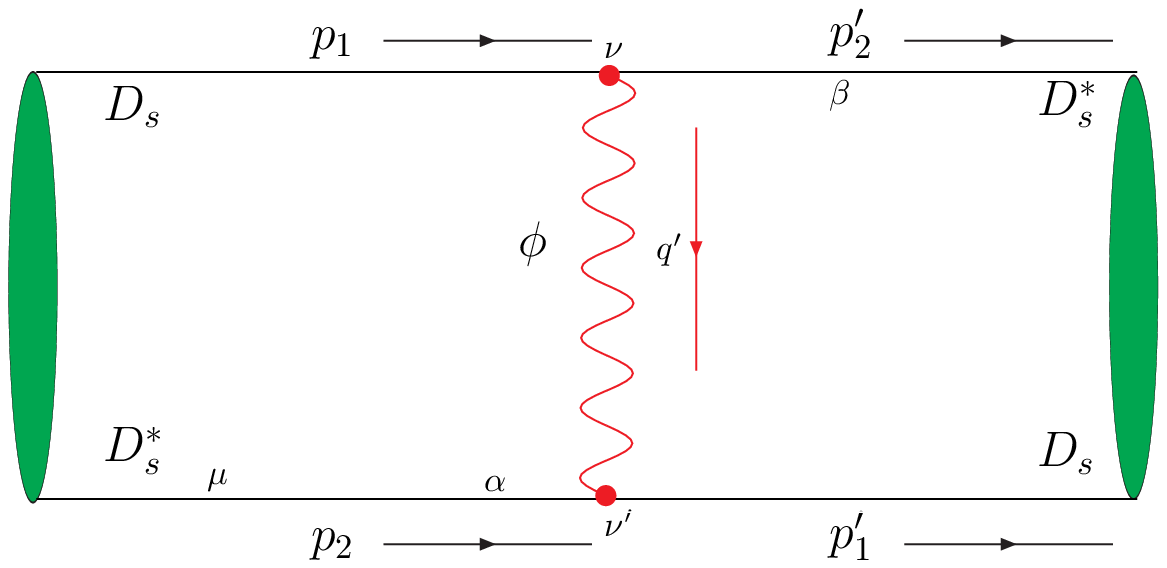}}\\
(a)\,\,\,\,\,\,\,\,\,\,\,\,\,\,\,\,\,\,\,\,\,\,\,\,\,\,\,\,\,\,\,\,\,\,\,\,
\,\,\,\,\,\,\,\,\,\,\,\,\,\,\,\,\,\,\,\,\,\,\,\,\,\,\,\,\,\,\,\,\,\,\,\,
\,\,\,\,\,\,\,\,\,\,\,\,\,\,\,\,\,\,\,\,\,\,\,\,\,\,\,\,\,\,\,\,\,\,\,\,
\,\,\,\,\,\,\,\,\,\,\,\,\,\,\, (b)
\end{tabular}
\caption{ The leading order interaction diagrams of a bound state composed of a pseudoscalar and a vector by
exchanging $\phi$. }\label{DM23}
\end{figure}
\end{center}

With the Feynman diagrams depicted in Fig. \ref{DM22} and Fig. \ref{DM23}, we eventually obtain
\begin{eqnarray}
K_{V\alpha\beta}(p,p')&&=K_{V1\alpha\beta}(p,p',m_\phi)+{
K_{V2\alpha\beta}}(p,p',m_\phi)+{
K_{V3\alpha\beta}}(p,p',m_\eta),\nonumber\\{K_{V1\alpha\beta}}(p,p',m_{
V})
&&=C_{V1}g_{_{DDV}}\{g_{_{D^*D^*V}}[g_{\alpha\beta}(p_2+p_2')\cdot(p_1+p_1')+g_{\alpha\beta}\frac{q\cdot(p_1+p_1')q\cdot(
p_2+p_2')}{m_{V}^2}] \nonumber\\&&+2g'_{_{D^*D^*V}}[q^\alpha
(p_1+p_1')^\beta-q^\beta
(p_1+p_1')^\alpha]\}\frac{i}{q^2-m_{V}^2}F(q)^2\nonumber\\{K_{V2\alpha\beta}}(p,p',m_{
V})&& = C_{V2}\varepsilon^{\mu\nu\beta\tau}(q'_\nu
g_{\lambda\mu}-q'_\mu
g_{\lambda\nu})(p_2'-p_1)_\tau\varepsilon^{\mu'\nu'\alpha\tau'}(q'_\mu
g_{\lambda'\nu}-q'_\nu
g_{\lambda'\mu})(p_2-p_1')_{\tau'}\nonumber\\&&\frac{i}{q'^2-m_{V}^2}(-g^{\lambda\lambda'}+q'^\lambda
q'^{\lambda'}/m^2_V)F(q')^2,\nonumber\\K_{V3\alpha\beta}(p,p',m_P)&&=C_{V3}g_{_{DD^*P}}^2\frac{i}{q^2-m_{P}^2}q_\alpha
q_\beta F(q)^2,
\end{eqnarray}
where $q'=p_1-p_2'$, $m_V$ represents the mass of vector meson (e.g., $\phi$), and $m_P$ represents mass of pseudoscalar meson (e.g., $\eta$), . The contributions from Fig. \ref{DM22} are
included in $K_{V3\alpha\beta}(p,p',m_\eta)$, and those from Fig.
\ref{DM23} (a) and (b) are included in
$K_{V1\alpha\beta}(p,p',m_\phi)$ and $K_{V2\alpha\beta}(p,p',m_\phi)$, respectively.
 The coefficients $C_{V1}$, $C_{V2}$ and $C_{V3}$ are 1, 1 and $\frac{2}{3}$,  respectively.

Defining $ K_{V}(p,p')={ K_{V\alpha\beta}}(q)(\frac{p_{2}^\mu
p_{2}^\alpha}{m_2^2}-g^{\mu\alpha})(\frac{P_\mu
P^\beta}{M^2}-g_\mu^\beta)$ and setting $p_0=q_0=0$, we derive the B-S
equation which is similar to Eq. (\ref{3-dim-BS1}) but possesses
a different kernel,
\begin{eqnarray} \label{3-dim-BS2}
{E^2-(E_1+E_2)^2\over (E_1+E_2)/E_1E_2}
\mathcal{\psi}_{{}_V}^{}({\bf p}) ={i\over
2}\int{d^3\mathbf{p}'\over(2\pi)^3}\, {\overline{} K_{V}}({\bf
p},{\bf p}')\mathcal{\psi}_{{}_V}^{}({\bf p}')\,,
\end{eqnarray}
where
\begin{eqnarray}
K_{V}(\mathbf{p},\mathbf{p}')&&=K_{V1}(\mathbf{p},\mathbf{p}',m_\phi)+{
K_{V2}}(\mathbf{p},\mathbf{p}',m_\phi)+{
K_{V3}}(\mathbf{p},\mathbf{p}',m_\eta),
\end{eqnarray}
where the expressions of $K_{V1}(\mathbf{p},\mathbf{p}',m_\phi)$,
$K_{V2}(\mathbf{p},\mathbf{p}',m_\phi)$ and ${
K_{V3}}(\mathbf{p},\mathbf{p}',m_\eta)$ can be found  in Ref. \cite{Ke:2021rxd}.

\begin{center}
\begin{figure}[htb]
\begin{tabular}{cc}
\scalebox{0.5}{\includegraphics{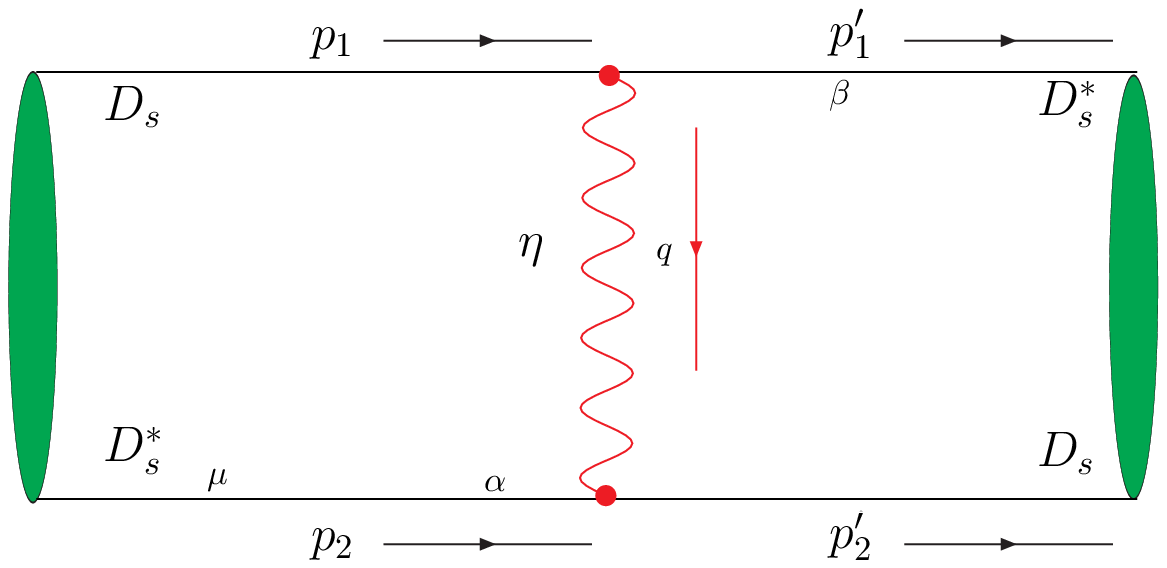}}\,\,\,\,\,\,\,\,\,\,\,\,\,\,\,\,\,\,\,\,\scalebox{0.5}{\includegraphics{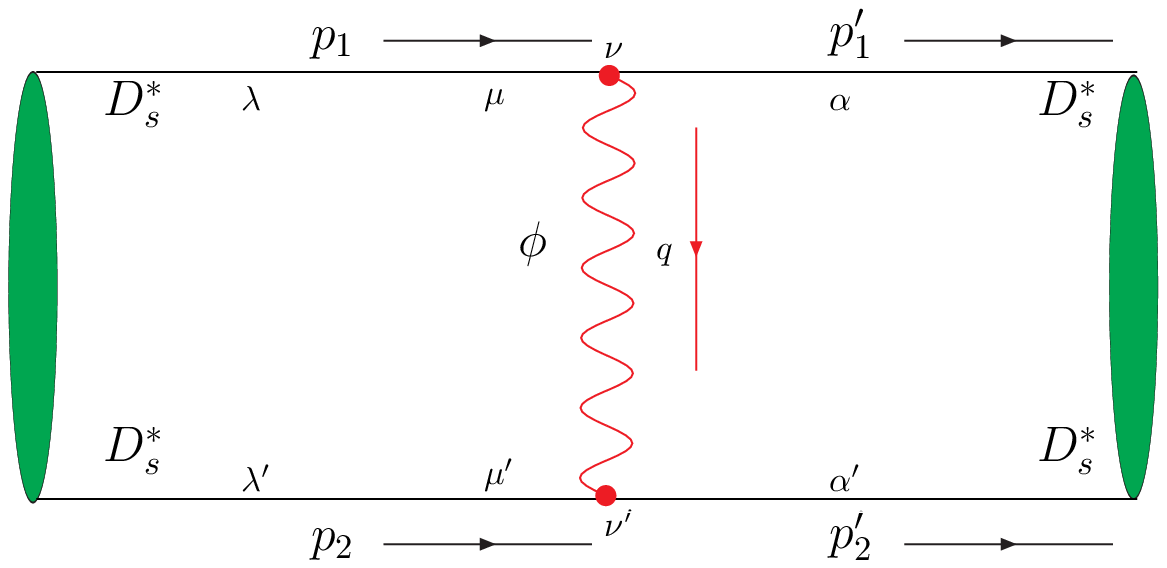}}
\\
(a)\,\,\,\,\,\,\,\,\,\,\,\,\,\,\,\,\,\,\,\,\,\,\,\,\,\,\,\,\,\,\,\,\,\,\,
\,\,\,\,\,\,\,\,\,\,\,\,\,\,\,\,\,\,\,\,\,\,\,\,\,\,\,\,\,\,\,\,\,\,\,\,
\,\,\,\,\,\,\,\,\,\,\,\,\,\,\,\,\,\,\,\,\,\,\,\,\,\,\, (b)
\end{tabular}
\caption{A bound state composed of two vectors. (a) $\eta$ is
exchanged. (b) $\phi$  is exchanged. }\label{DM24}
\end{figure}
\end{center}

\subsection{The bound state ($0^+$) composed of two vector mesons}

The quantum number $J^P$ of the bound state composed of two
vector mesons can only be $0^+$ or $2^{+}$ since the total wave function should be symmetric. The B-S wave
function of $0^+$ state $|{S'}\rangle$ is defined as following:
\begin{eqnarray}\label{definition-BS}  \langle 0 | {\rm
T}\,\phi_{1}^\mu(x_1)\phi^\nu_2(x_2) |{S'} \rangle =
{\chi}_{{}_{{0}}}^{}(x_1,x_2)g^{\mu\nu}\,.
\end{eqnarray}
The equation for the B-S wave function is given by
\begin{eqnarray} \label{4-dim-BS21}
\chi_{_{{}_{0}}}({
p})=\frac{1}{4}\Delta_{1\mu\lambda}\int{d^4{p}'\over(2\pi)^4}\,K_0^{\alpha\alpha'\mu\mu'}({
p},{ p}')\chi_{_{{}_{0}}}^{}({
p}')\Delta_{2\mu'\lambda'}g_{\alpha\alpha'}g^{\lambda\lambda'},
\end{eqnarray}
where
$\Delta_{j\mu\lambda}=\frac{i}{p_j^2-m_j^2}(\frac{p_{j\mu}p_{j\lambda}}{m_2^2}-g_{\mu\lambda})$.

With the effective interaction Feynman diagrams depicted in Fig. \ref{DM24}, we obtain
\begin{eqnarray}\label{k0}
  K_{0}^{\alpha\alpha'\mu\mu'}(p,p')&&=\sqrt{2}K_{01}^{\alpha\alpha'\mu\mu'}(p,p',m_\phi)+\sqrt{2}
  K_{03}^{\alpha\alpha'\mu\mu'}(p,p',m_\eta),\nonumber\\K_{01}^{\alpha\alpha'\mu\mu'}(p,p',m_V)&&=iC_{01}\frac{{q^\nu
q^{\nu'}}/{{m_V}^2}-g^{\nu{\nu'}}}{q^2-m_{V}^2}[g_{_{D^*D^*V}}g^{\alpha\mu}
(p_1+p_1')_\nu-2g_{_{D^*D^*V}}'({q}^{\alpha} g_{\mu\nu}-{q}^\mu
g_{\alpha\nu})]\nonumber\\&&[g_{_{D^*D^*V}}g^{\alpha'\mu'}
(p_2+p_2')_{\nu'}+2g_{_{D^*D^*V}}'({q}^{\alpha'}
g_{\mu'\nu'}-{q}^{\mu'} g_{\alpha'\nu'})]F(q)^2,\nonumber\\K_{03}^{\alpha\alpha'\mu\mu'}(p,p',m_P)&&=C_{03}g_{_{D^*D^*P}}^2\varepsilon^{\alpha\beta\mu\nu}q_\nu
(p_1+p_1')_\beta\varepsilon^{\alpha'\beta'\mu'\nu'}q_{\nu'}
(p_2+p_2')_{\beta'}\frac{-i}{q^2-M_\pi^2}F(q)^2.
\end{eqnarray}
The contribution from vector-meson-exchange ($\phi$) is included in
$K_{01}^{\alpha\alpha'\mu\mu'}(p,p',m_V)$ and that for exchanging
pseudoscalar ($\eta$) is included in
$K_{03}^{\alpha\alpha'\mu\mu'}(p,p',m_P)$ (the label $K_{01,03}$ are based on the definition of functions in Ref. \cite{Ke:2021rxd} ). The coefficients $C_{01}$ and $C_{03}$ are $1$ and $\frac{2}{3}$, respectively.

Defining $ K_{0}(p,p')=\frac{1}{4}{ K_0^{\alpha\alpha'\mu\mu'}}({
p},{ p}')(\frac{p_{2\mu'}
p_{2\lambda'}}{m_2^2}-g_{\mu'\lambda'})(\frac{{p_{1}}_\mu
{p_{1}}_{\lambda}}{m_1^2}-g_{\mu\lambda})$, we derive the B-S equation which is similar to Eqs. (\ref{3-dim-BS1}) but possesses
a different kernel.

The B-S equation can be reduced to
\begin{eqnarray} \label{3-dim-BS3}
{E^2-(E_1+E_2)^2\over (E_1+E_2)/E_1E_2}
\mathcal{\psi}_{{}_0}^{}({\bf p}) ={i\over
2}\int{d^3\mathbf{p}'\over(2\pi)^3}\, {\overline{} K_{{0}}}({\bf
p},{\bf p}')\mathcal{\psi}_{{}_0}^{}({\bf p}')\,,
\end{eqnarray}
where
\begin{eqnarray}
K_{0}(\mathbf{p},\mathbf{p}')&&=\sqrt{2}K_{01}(\mathbf{p},\mathbf{p}',m_\phi)+\sqrt{2}{
K_{03}}(\mathbf{p},\mathbf{p}',m_\eta).
\end{eqnarray}
The expressions of $K_{01}(\mathbf{p},\mathbf{p}',m_V)$ and ${
K_{03}}(\mathbf{p},\mathbf{p}',m_P)$ also can be found  in Ref. \cite{Ke:2021rxd}.

\subsection{The B-S equation of $2^+$ state $|T'\rangle$ which is composed of two vectors}
\label{two_v}
 The B-S wave-function of $2^+$ state composed of two axial-vectors is written as
 \begin{eqnarray} \label{4-dim-BS23}
\langle0|T\phi^\alpha(x_1)\phi^{\alpha'}(x_2)|T'\rangle=\frac{1}{\sqrt{5}}\chi_{_{2}}(x_1,x_2)\varepsilon^{\alpha\alpha'}
,
\end{eqnarray}
where $\varepsilon^{\alpha\alpha'}$ is the polarization vector of the $2^+$ state. Summing over the polarizations of $\varepsilon^{\alpha\alpha'}\varepsilon^{*\beta\beta'}$ one can obtain $(Q^{\alpha\beta}Q^{\alpha'\beta'}+Q^{\alpha\beta'}Q^{\alpha'\beta})/2-Q^{\alpha\alpha'}Q^{\beta\beta'}/3$ with $Q^{\alpha\beta}=P^\alpha P^\beta/M^2-g^{\alpha\beta}$.

The B-S equation can be expressed as
\begin{eqnarray} \label{4-dim-BS43}
\chi_{_{2}}(p)
=\frac{1}{5}\varepsilon^{\lambda\lambda'}\Delta_{1\mu\lambda}\int{d^4{q}\over(2\pi)^4}\,K_2^{\alpha\alpha'\mu\mu'}(p,p')
\varepsilon_{\alpha\alpha'}\chi_{_{2}}(q)\Delta_{2\mu'\lambda'}\,,
\end{eqnarray}
where $K_2^{\alpha\alpha'\mu\mu'}(p,p')$ is the same as $K_0^{\alpha\alpha'\mu\mu'}(p,p')$ in Eq. (\ref{k0}).

Defining $K_{2}(p,p')= \frac{{
K_2^{\alpha\alpha'\mu\mu'}}({ p},{
p}')}{5}\varepsilon^{\lambda \lambda'
}(\frac{p_{2\mu'}
p_{2\lambda'}}{m_2^2}-g_{\mu'\lambda'})(\frac{{p_{1}}_\mu
{p_{1}}_{\lambda}}{m_1^2}-g_{\mu\lambda})\varepsilon_{\alpha\alpha'}$, we find reduce the
 B-S equation to the following form
\begin{eqnarray} \label{3-dim-BS5}
{E^2-(E_1+E_2)^2\over (E_1+E_2)/E_1E_2}
\mathcal{\psi}_{{}_{2}}^{}({\bf p}) ={i\over
2}\int{d^3\mathbf{p}'\over(2\pi)^3}\, {\overline{} K_{2}}({\bf
p},{\bf p}')\mathcal{\psi}_{{}_{2}}^{}({\bf p}')\,,
\end{eqnarray}
where
\begin{eqnarray}
K_{2}(\mathbf{p},\mathbf{p}')&&=\sqrt{2}K_{21}(\mathbf{p},\mathbf{p}',m_\phi)+\sqrt{2}{
K_{23}}(\mathbf{p},\mathbf{p}',m_\eta).
\end{eqnarray}
The expressions of $K_{21}(\mathbf{p},\mathbf{p}',m_\phi)$ and ${
K_{23}}(\mathbf{p},\mathbf{p}',m_\eta)$ are presented in Ref. \cite{Ke:2021rxd}.

\section{Numerical results}
In this section, we solve the B-S equations  (\ref{3-dim-BS1}),
(\ref{3-dim-BS2}), (\ref{3-dim-BS3}) and
(\ref{3-dim-BS5}) to study whether these bound states can exist. Since we only focus on the ground state of a bound state, the function $\psi_{J}(\mathbf{p})$ ($J$ represents $S, V, 0$ or
$2$) only depends on
the norm of the three-momentum. Therefore, we can first integrate over the
azimuthal angle of the function in (\ref{3-dim-BS1}),
(\ref{3-dim-BS2}), (\ref{3-dim-BS3}) or
(\ref{3-dim-BS5})
$$\frac{i}{2}\int{d^3\mathbf{p}'\over(2\pi)^3}\, {\overline{} K_J}({\bf p},{\bf
p}')  $$  to obtain a potential form
$U_J(|\mathbf{p}|,|\mathbf{p}'|)$. After that,  the B-S equation turns into a one-dimension integral
equation
\begin{eqnarray} \label{3-dim-BS6}
\psi_J({\bf |p|}) ={(E_1+E_2)/E_1E_2\over E^2-(E_1+E_2)^2 }\int{d
\mathbf{|p}'|}\, {\overline{} U_J}({\bf |p|},{\bf
|p}'|)\psi_J({\bf |p}'|) .
\end{eqnarray}
When the potential
$U_{J}(\mathbf{p},\mathbf{p}')$  is attractive and strong enough,
the corresponding B-S equation has one or multiple solutions and we can obtain
the spectrum (or spectra) of the possible bound state(s).
In general, the standard way of solving such integral equation is to
discretize the variable and then perform algebraic operations. The detail of this approach can be found in \cite{Ke:2021rxd,Ke:2020eba,Ke:2012gm}.

In our calculation the values of the parameters $g_{_{D_sD_sV}},
g_{_{D_sD^*_sP}}, g_{_{D_sD^*_sV}}$, $g_{_{D^*_sD^*_sV}}$ and
$g'_{_{D^*_sD^*_sV}}$ are presented in Appendix A.
In Ref.\cite{Ke:2021rxd} which suggested $T^+_{cc}$ is a $D^0D^{*+}$ molecular state,
 $\Lambda$ was fixed to be $1.134$ GeV. For $D^{(*)}_sD^{(*)}_s$ and $B^{(*)}_sB^{(*)}_s$  system, we adjust $\Lambda$ around that value. The masses of the concerned constituent mesons $m_{D_s}$, $m_{D^*_s}$, $m_{B_s}$ and $m_{B^*_s}$ are directly taken from booklet of particle data group \cite{PDG10}.

\subsection{The results of $D^{(*)}_sD^{(*)}_s$ system}

Here we calculate the eigenvalues of bound states $D_s
D_s(0^+)$, $D_s D^{*}_s(1^+)$, $D^{*}_s  D^{*}_s(0^+)$ and $D^{*}_s
D^{*}_s(2^+)$, respectively. For fixed parameters $\Lambda=$ $1.134$ GeV and above mentioned coupling constants, all B-S equations are unsolvable, so we try to vary the parameter $\Lambda$ or coupling constants to search the solutions of these equations.  For  the state  of
$D_sD^{*}_s$ system we can obtain a solution with the binding energy $\Delta E=1$ MeV  when we set $\Lambda=2.564$ GeV or coupling constants to be 3.261 times the original. It implies the effective interaction between the two constituents is relatively weak. In table. \ref{tab:ev1} we list the values of $\Lambda$ for different $D^{(*)}D^{(*)}$ systems when  the binding energy $\Delta E$ is $1$ MeV. One can find the value of $\Lambda$ for the $D^{*}_sD^{*}_s$  with $J^p=0^+$ is closest to the value 1.134 GeV that we fixed in Ref. \cite{Ke:2021rxd}. We vary the binding energy of the $D^{*}_sD^{*}_s$ system from 0.1 MeV to 2 MeV and the corresponding values of $\Lambda$ are collected in table \ref{tab:ev1p} where one can find $\Lambda$ is not sensitive to the change of the binding energy.

In Refs. \cite{Cheng:2004ru,Meng:2007tk} the authors
suggested a relation: $\Lambda=m+\alpha \Lambda_{QCD}$ where $m$
is the mass of the exchanged meson, $\alpha$ is a number of $O(1)$
and $\Lambda_{QCD}=220$ MeV. According to this relation, the value of $\Lambda$ for $D^{*}_sD^{*}_s$ ($0^+$)  locates at the upper limit, hence the loose bound state composed of $D^{*}_sD^{*}_s$ ($0^+$) exist with a very low probability.

\begin{table}
\caption{ The $\Lambda$ for the ground $D^{(*)}_sD^{(*)}_s$ systems with fixed binding energy $\Delta E=1$ MeV (in
unit of GeV).} \label{tab:ev1}
\begin{tabular}{c|c|c|c|c|c}\hline\hline
 ~~~~~~~~   &  ~~~$D_s D_s(0^+)$~~~    &
 ~~~$D_s D^{*}_s(1^+)$~~~&  ~~~$D^{*}_sD^{*}_s(0^+)$~~~&   ~~~$D^{*}_sD^{*}_s(2^+)$~~~   \\\hline
 $\Lambda$    &2.848   & 2.564  &  1.519 &13.78 \\
\hline\hline
\end{tabular}
\end{table}

\begin{table}
\caption{ The values of $\Lambda$ for the ground $D^{*}_sD^{*}_s (0^+)$ system with different binding energies.} \label{tab:ev1p}
\begin{tabular}{c|c|c|c|c|c|c}\hline\hline
 $\Delta E$ (MeV) &0.1  &0.2       & 0.5  &  1 &1.5&2 \\\hline
 $\Lambda$ (GeV) &1.501  &1.502      & 1.506  &  1.511 &1.516&1.519 \\
\hline\hline
\end{tabular}
\end{table}

\subsection{The results of the $B^{(*)}_sB^{(*)}_s$ system}

Under flavor $SU(3)$ symmetry and heavy quark limit, $g_{_{B_sB_sV}}$,
$g_{_{B_sB^{*}_sP}}$, $g_{_{B_sB^{*}_sV}}$,
$g_{_{B^{*}_sB^{*}_sV}}$ and $g'_{_{B^{*}_sB^{*}_sV}}$ are approximately equal to $g_{_{D_sD_sV}}$, $g_{_{D_sD^{*}_sP}}$, $g_{_{D_sD^{*}_sV}}$, $g_{_{D^{*}_sD^{*}_sV}}$
, $g'_{_{D^{*}_sD^{*}_sV}}$, respectively.

We set the binding energy to be 1 MeV and vary the parameters $\Lambda$  for the $B_s^{(*)}B_s^{(*)}$ systems which are shown in table
\ref{tab:ev2}. We
find that  the value of $\Lambda$ fixed from the state of $B_s^{(*)}B_s^{(*)}$ is closer to 1.134 GeV than that of  the corresponding
state of $D_s^{(*)}D_s^{(*)}$.
%
%
%
%
Especially the $\Lambda$ value of the $B_s^{*} B_s^{*}$ system with $0^+$ is  a little larger than 1.134 GeV, which means that two $B_s^{*}$ mesons perhaps form a loose bound state.

\begin{table}
\caption{  $\Lambda$ for the ground $B^{(*)}_sB^{(*)}_s$ system with fixed binding energy $\Delta E=1$ MeV (in
unit of GeV).} \label{tab:ev2}
\begin{tabular}{c|c|c|c|c|c}\hline\hline
 ~~~~~~~~   &  ~~~$B_s B_s(0^+)$~~~   &
 ~~~$B_s B^{*}_s(1^+)$~~~&  ~~~$B^{*}_sB^{*}_s(0^+)$~~~&   ~~~$B^{*}_sB^{*}_s(2^+)$~~~   \\\hline
 $\Lambda$    &1.677      & 1.653  &  1.293 &2.231 \\
\hline\hline
\end{tabular}
\end{table}

\begin{table}
\caption{ The value of $\Lambda$ for the ground $B^{*}_sB^{*}_s (0^+)$ system with varying binding energies $\Delta E$ .} \label{tab:ev2p}
\begin{tabular}{c|c|c|c|c|c|c}\hline\hline
 $\Delta E$ (MeV) &0.1  &0.2       & 0.5  &  1 &1.5&2 \\\hline
 $\Lambda$ (GeV) &1.273  &1.276    & 1.284 &  1.293 &1.301&1.308 \\
\hline\hline
\end{tabular}
\end{table}

\section{A brief summary}
Following the approach in our recent work we study whether two $D^{(*)}_s$ or $B^{(*)}_s$ mesons can
form a hadronic molecule within the B-S framework. In Ref.
\cite{Guo:2007mm,Feng:2011zzb,Ke:2012gm,Ke:2020eba,Ke:2019bkf} the
B-S equations were applied to the systems of one particle and one antiparticle. In Ref.\cite{Ke:2021rxd} we studied the possible bound states of particle-particle system (two $D^{(*)}$ or $B^{(*)}$ mesons).  In this work we extend this framework to the $D^{(*)}_s$ or $B^{(*)}_s$ systems and compute binding energy of such hadronic molecules by solving B-S equations.

We employ heavy meson chiral perturbation theory and leading order (single-meson exchange) approximation to calculate the interaction kernels of the B-S equations for the $D^{(*)}_s$ or $B^{(*)}_s$ systems, where $\eta$ or $\phi$ is exchanged. All coupling constants are taken from relevant references. Adopting the fixed value of $\Lambda$  from the analysis of $T_{cc}^+$ under the hypothesis that
$T_{cc}^+$ is a bound state of $D^0D^{*+}$ with $I=0$ and $J=1$ we find all these B-S equations have no solution. Then we
set the binding energy $\Delta E=1$ MeV and test different values of $\Lambda$ for  two $D^{(*)}_s$ or $B^{(*)}_s$ systems with defined quantum number.
We find for most states, a larger $\Lambda$ or coupling constants are needed to form bound states.
It is noted that the $\Lambda$ values of   $B^{*}_sB^{*}_s$ with $J^P=0^+$ is close to that fixed from $T_{cc}^+$,  which suggests that such a molecular bound state may exist. Since in our calculations, the values of model parameters are estimated based on previous literature which span a relatively large range, we cannot expect all the numerical results to be very accurate. However, the major goal of this work is to qualitatively analyze whether two $D^{(*)}_s$ or $B^{(*)}_s$ mesons can form a molecular state. Even if the numerical estimation is not precise, our result can still provide a useful guidance for the future study of double charmed hardonic molecular state. Further theoretical and experimental works are needed for gaining a better understanding of the double charmed exotic states.

\section*{Acknowledgments}

 This work is supported by the National Natural Science Foundation
of China (NNSFC) under the contract No. 12075167.
\appendix

\section{The effective interactions}
The effective interactions can be found
in\cite{Colangelo:2005gb,Colangelo:2012xi,Ding:2008gr}
\begin{eqnarray}
&&\mathcal{L}_{_{DDV}}=g_{_{DDV}}(D_{b}\stackrel{\leftrightarrow}{\partial}_{\beta}
D^{\dag}_{a})(
\mathcal{V}^\beta)_{ba},\\&&\mathcal{L}_{_{DD^*V}}=ig_{_{DD^*V}}\varepsilon^{\alpha\beta\mu\nu}({\partial}_{\alpha}\mathcal{V}_\beta
-{\partial}_{\beta}\mathcal{V}_\alpha)_{ba}(\partial_\nu
D_{b}D^{*\mu\dag}_{a}-\partial_\nu
D^{*\mu\dag}_{b}D_{a}),\\&&\mathcal{L}_{_{DD^*P}}=g_{_{DD^*P}}D_{b}(\partial_\mu
\mathcal{M})_{ba}D^{*\mu\dag}_{a}+g_{_{DD^*P}}D^{*\mu}_{b}(\partial_\mu
\mathcal{M})_{ba}D^{\dag}_{a},\\&&\mathcal{L}_{_{D^*D^*P}}=g_{_{D^*D^*P}}(D^{*\mu}_{b}\stackrel{\leftrightarrow}{\partial}^{\beta}
D^{*\alpha\dag}_{a})(\partial^\nu
\mathcal{M})_{ba}\varepsilon_{\nu\mu\alpha\beta},
\\&&\mathcal{L}_{_{D^*D^*V}}=ig_{_{D^*D^*V}}(D^{*\nu}_{b}\stackrel{\leftrightarrow}{\partial}_{\mu}
D^{*\dag}_{a\nu})(
\mathcal{V})_{ba}^\mu+ig'_{_{D^*D^*V}}(D^{*\mu}_{b}
D^{*\nu\dag}_{a}-D^{*\mu\dag}_{b} D^{*\nu}_{a})(
\partial_\mu\mathcal{V}_\nu-\partial_\nu\mathcal{V}_\mu)_{ba}
\end{eqnarray}
where $a$ and $b$ represent index of SU(3) flavor group for three light quarks. In
Refs.\cite{Ding:2008gr} $\mathcal{M}$ and $\mathcal{V}$ are
$3\times 3$ hermitian and traceless matrixs $
\left(\begin{array}{ccc}
        \frac{\pi^0}{\sqrt{2}}+\frac{\eta}{\sqrt{6}} &\pi^+ &K^+ \\
         \pi^- & -\frac{\pi^0}{\sqrt{2}}+\frac{\eta}{\sqrt{6}}&K^0\\
         K^-& \bar{K^0} & -\sqrt{\frac{2}{3}}\eta
      \end{array}\right)$
       and $
\left(\begin{array}{ccc}
        \frac{\rho^0}{\sqrt{2}}+\frac{\omega}{\sqrt{2}} &\rho^+ &K^{*+} \\
         \rho^- & -\frac{\rho^0}{\sqrt{2}}+\frac{\omega}{\sqrt{2}}&K^{*0}\\
         K^{*-}& \bar{K^{*0}} & \phi
      \end{array}\right)$, respectively.

In the flavor $SU(3)$ symmetry and heavy quark limit, the above coupling constants
are given by $g_{_{D_sD_sV}}=\frac{\beta g_V}{\sqrt{2}},
g_{_{D_sD^*_sV}}=\frac{\lambda g_V}{\sqrt{2}},
g_{_{D^*_sD^*_sP}}=\frac{g}{f_\pi},$ $g_{_{D_sD^*_sP}}=g_{_{DD^*P}}=-\frac{2g}{f_{\pi}}
\sqrt{M_{D}M_{D^*}}, g_{_{D^*_sD^*_sV}}=-\frac{\beta
g_V}{\sqrt{2}},\,\, g'_{_{D^*_sD^*_sV}}=g'_{_{D^*D^*V}}=-\sqrt{2}\lambda g_V M_{D^*}$
with $f_\pi=132$ MeV\cite{Colangelo:2005gb},
$g=0.64$\cite{Colangelo:2012xi}, $\kappa=g$, $\beta=0.9$,
$g_V=5.9$\cite{Falk:1992cx} and $\lambda =0.56$ GeV$^{-1}$\cite{Chen:2019asm}.

\end{document}